\documentclass[twocolumn,showpacs,showkeys,amsmath,amssymb,superscriptaddress,floatfix,nofootinbib]{revtex4}

\usepackage{graphicx}
\usepackage{bm} 
\usepackage{subfigure}
\def\vec#1{\mathchoice{\mbox{\boldmath$\displaystyle#1$}}
{\mbox{\boldmath$\textstyle#1$}}
{\mbox{\boldmath$\scriptstyle#1$}}
{\mbox{\boldmath$\scriptscriptstyle#1$}}}
\makeatletter
\newcommand\erfc{\mathop{\operator@font erfc}\nolimits}
\def\slashchar#1{\setbox0=\hbox{$#1$}
   \dimen0=\wd0 \setbox1=\hbox{/} \dimen1=\wd1
   \ifdim\dimen0>\dimen1 \rlap{\hbox to \dimen0{\hfil/\hfil}} #1
   \else  \rlap{\hbox to \dimen1{\hfil$#1$\hfil}} / \fi}

\begin{document}
 
\title{
Dynamics of anisotropic plasma at the early stages of relativistic heavy-ion collisions
\footnote{Supported in part by the Polish Ministry of Science and Higher Education, grant  N202 034 32/0918.}}

\author{Wojciech Florkowski} 
\email{Wojciech.Florkowski@ifj.edu.pl}
\affiliation{The H. Niewodnicza\'nski Institute of Nuclear Physics, Polish Academy of Sciences, PL-31342 Krak\'ow, Poland}
\affiliation{Institute of Physics, Jan Kochanowski University, PL-25406~Kielce, Poland} 

\author{Radoslaw Ryblewski} 
\email{Radoslaw.Ryblewski@ifj.edu.pl}
\affiliation{The H. Niewodnicza\'nski Institute of Nuclear Physics, Polish Academy of Sciences, PL-31342 Krak\'ow, Poland}

\date{January 26, 2008}

\begin{abstract}
We discuss the evolution of anisotropic boost-invariant quark-gluon plasma possibly created at the early stages of relativistic heavy-ion collisions. Our considerations are based on the recently proposed formalism that is an extension of the relativistic perfect-fluid hydrodynamics. We analyze i) the pure partonic system described by the anisotropic phase-space distribution function and ii) the system of partons interacting with the local magnetic fields. The second analysis is a simplified attempt to include the effects of color fields on the particle dynamics. Our model results are discussed in the context of early thermalization and isotropization.  Under general assumptions of the particle, energy, and momentum conservations we show that for large evolution times the ratio of the longitudinal and transverse pressures of the pure partonic system tends to zero. Hence, the system with the initial momentum distribution elongated along the beam axis always passes through the isotropic stage where the transverse and longitudinal pressures are equal. The inclusion of the magnetic field in this case gives negative contribution to the longitudinal pressure, hence the transient stage when the total longitudinal and transverse pressures become equal may be reached earlier, depending on the strength of the field.  
\end{abstract}



\maketitle 

\section{Introduction}
\label{sect:intro}

The data collected in the heavy-ion experiments at the Relativistic Heavy-Ion Collider (RHIC) are most commonly interpreted as the evidence that the matter produced in relativistic heavy-ion collisions equilibrates very fast (presumably within a fraction of 1~fm/c) and its behavior is very well described by the perfect-fluid hydrodynamics  \cite{Kolb:2003dz,Huovinen:2003fa,Shuryak:2004cy,Teaney:2001av,Hama:2005dz,Hirano:2007xd,Nonaka:2006yn}. Such features are naturally explained by the assumption that the produced matter is a strongly coupled quark-gluon plasma (sQGP) \cite{Shuryak:2004kh}. Another explanation assumes that the plasma is weakly interacting, however the plasma instabilities lead to the fast isotropization of matter, which in turn helps to achieve equilibration \cite{Mrowczynski:2005ki}. Recently, it has been also shown that the model assuming thermalization of the transverse degrees of freedom only \cite{Bialas:2007gn} is consistent with the data describing the transverse-momentum spectra and the elliptic flow coefficient $v_2$. This result indicates  that the assumption of the fast thermalization/isotropization might be relaxed. 

In view of the problems related to the thermalization and isotropization of the plasma, it is useful to develop and analyze the models which can describe anisotropic systems. Recently, an effective model describing the anisotropic fluid/plasma dynamics in the early stages of relativistic heavy-ion collisions has been introduced  \cite{Florkowski:2008ag}. The model has the structure similar to the perfect-fluid relativistic hydrodynamics -- the equations of motion follow from the conservation laws. However, it admits the possibility that the longitudinal and transverse pressures are different (as usual, the longitudinal direction is defined here by the beam axis). The main characteristic feature of the proposed model is the use of the pressure relaxation function $R$ which determines the time changes of the ratio of the transverse and longitudinal pressures and, possibly, defines the way how the system becomes isotropic, i.e., how the transverse and longitudinal pressures become equilibrated. The role of the pressure relaxation function is similar and complementary to the role played by the equation of state. It characterizes the material properties of the medium whose spacetime dynamics is otherwise governed  by the conservation laws. 

In this paper we develop the formulation of Ref. \cite{Florkowski:2008ag} in two ways: i) we introduce  the microscopic interpretation for the relaxation function in the case where the considered system consists of particles whose behavior is described by the momentum-anisotropic phase-space distribution function, ii) we include the effects of the fields considering the case of anisotropic magnetohydrodynamics -- this approach may be regarded as a very crude attempt to include the effects of color fields on the particle dynamics.

Our first finding is that the use of the anisotropic phase-space distribution functions leads inevitably to the pressure relaxation functions $R$ which imply that the ratio of the longitudinal and transverse pressures tends asymptotically to zero, $P_L/P_T \to 0$. This behavior is complementary to the recent results obtained from the analyses of the early-stage partonic free-streaming \cite{Jas:2007rw,Broniowski:2008qk}. In fact, our approach includes the free-streaming as the special case, however, it may be also applied in the cases where the collisions are present but their effect does not change the assumed generic form of the phase-space distribution function.  The time asymptotics $P_L/P_T \to 0$ means that the systems with the initial prolate momentum shape, i.e., the systems that are initially elongated along the beam axis in the momentum space with \mbox{$P_L > P_T$} (see for example Refs. \cite{Jas:2007rw,Randrup:2003cw}), naturally pass through the transient isotropic stage where the transverse and longitudinal pressures are equal. Our second finding follows from the study of the magnetohydrodynamic model. We show that the inclusion of the fields lowers the longitudinal pressure and increases the transverse pressure, hence, for the initially prolate systems the stage when the {\it total} longitudinal and transverse pressures become equal may be reached earlier, depending on the strength of the field.  

The two models discussed by us cannot explain the phenomenon of reaching the {\it stable} isotropic stage. However, we indicate that the presence of the fields may have an impact on the process of isotropization, presumably restored by the effects of those collisions and/or field instabilities that are not taken into account in the present formalism. From the practical point of view, our formalism allows for the determination of the space-time evolution of the color-neutral anisotropic distributions, which may be used, for example, as the background distributions in the analysis of the plasma instabilities. Interestingly, in the case where we have initially $P_L > P_T$, the dynamics of the background and the plasma instable behavior evolve in the same direction, i.e., the two processes restore the equality of pressures. On the other hand, if the initial configuration has $P_T > P_L$, the plasma instabilities must compete with the growing asymmetry of the background (see Ref. \cite{Rebhan:2008uj,Rebhan:2008ry} where the growth of the instabilities was studied in the Bjorken longitudinal expansion and substantially large times of about 20 fm/c were found for RHIC). The interplay of such competing processes may be directly related to the problem of very fast thermalization/isotropization taking place in relativistic heavy-ion collisions. In addition, the discussed by us transformation of the longitudinal pressure into the transverse one is an interesting phenomenon analyzed in the context of the RHIC HBT puzzle -- we note that the recently proposed explanations of this puzzle suggest a very fast formation of the transverse flow \cite{Gyulassy:2007zz,Broniowski:2008vp,Pratt:2008bc}. 

The paper is organized as follows: In Sect. II we consider the anisotropic system of partons described by the momentum-anisotropic phase-space distribution function. We calculate the moments of the distribution function, determine the pressure relaxation function $R$, and argue that the general form of $R$ implies that the ratio of the longitudinal and transverse pressures tends asymptotically to zero.  In Section III we consider an example of boost-invariant magnetohydrodynamics. We analyze in detail the consistency of this approach and show that it may be treated as the special case of the formalism developed in \cite{Florkowski:2008ag} with the appropriate relaxation function $\hat R$. We also show how the presence of the local magnetic fields affects the dynamics of particles. We conclude in Sect. IV.

Below we assume that particles are massless and we use the following definitions for rapidity and spacetime rapidity,
\begin{eqnarray}
y = \frac{1}{2} \ln \frac{E_p+p_\parallel}{E_p-p_\parallel}, \quad
\eta = \frac{1}{2} \ln \frac{t+z}{t-z}, \label{yandeta} 
\end{eqnarray}
which come from the standard parameterization of the four-momentum and spacetime coordinate of a particle,
\begin{eqnarray}
p^\mu &=& \left(E_p, {\vec p}_\perp, p_\parallel \right) =
\left(p_\perp \cosh y, {\vec p}_\perp, p_\perp \sinh y \right), \nonumber \\
x^\mu &=& \left( t, {\vec x}_\perp, z \right) =
\left(\tau \cosh \eta, {\vec x}_\perp, \tau \sinh \eta \right). \label{pandx}
\end{eqnarray} 
Here the quantity $p_\perp$ is the transverse momentum
\begin{equation}
p_\perp = \sqrt{p_x^2 + p_y^2},
\label{energy}
\end{equation}
and $\tau$ is the (longitudinal) proper time
\begin{equation}
\tau = \sqrt{t^2 - z^2}.
\label{tau}
\end{equation} 
Throughout the paper we use the natural units where $c=1$ and $\hbar=1$.

\section{Anisotropic system of particles}
\label{sect:aniso-system}

In this Section we consider a system of particles/partons described by the distribution function which is asymmetric in the momentum space, i.e., its dependence on the longitudinal and transverse momentum is different. We calculate the particle current, the energy-momentum tensor, and the entropy current of such a system. As the special case we consider the exponential Boltzmann-like distributions frequently used in other studies. This Section is also used to introduce general concepts of anisotropic plasma dynamics, which will be applied to the system consisting of particles and fields in the next Section. 

\subsection{Anisotropic momentum distribution}
\label{sect:aniso-distribution}

We take into account the phase space distribution function whose dependence on the transverse and longitudinal momentum is determined by the two space-time dependent scales, $\lambda_\perp$ and $\lambda_\parallel$, namely
\begin{equation}
f = f\left( \frac{p_\perp}{\lambda_\perp},\frac{|p_\parallel|}{\lambda_\parallel}\right).
\label{Fxp1}
\end{equation}
The form (\ref{Fxp1}) is valid in the local rest frame of the plasma element. For boost-invariant systems, the explicitly covariant form of the distribution function has the structure
\begin{equation}
f  = f\left( \frac{\sqrt{(p \cdot U)^2 - (p \cdot V)^2 }}{\lambda _\perp }, 
\frac{|p \cdot V|}{\lambda _\parallel  }\right),
\label{Fxp2}
\end{equation}
where
\begin{equation}
U^{\mu} = ( u_0 \cosh\eta,u_x,u_y, u_0 \sinh\eta),
\label{U}
\end{equation}
\begin{equation}
V^{\mu} = (\sinh\eta,0,0,\cosh\eta),
\label{V}
\end{equation}
and $u^0, u_x, u_y$ are the components of the four vector
\begin{equation}
u^\mu = \left(u^0, {\vec u}_\perp, 0 \right) = \left(u^0, u_x, u_y, 0 \right).
\label{smallu}
\end{equation}
The four-velocity $u^\mu$ is normalized to unity
\begin{equation}
 u^\mu u_\mu = u_0^2 - u_x^2 - u_y^2 = 1.
\label{normsmallu}
\end{equation}
The four-vector $U^\mu$ describes the four-velocity of the plasma element. It may be obtained from $u^\mu$ by the Lorentz boost along the $z$ axis with rapidity $\eta$. The appearance of the four-vector $V^\mu$  is a new feature related to the anisotropy -- in the rest frame of the plasma element we have $V^\mu = (0,0,0,1)$. We note that the four vectors $U^\mu$ and $V^\mu$ satisfy the following normalization conditions:
\begin{equation}
U^\mu U_\mu = 1, \quad V^\mu V_\mu = -1, \quad U^\mu V_\mu = 0.
\label{normort}
\end{equation}

The boost-invariant character of Eq. (\ref{Fxp2}) is immediately seen if we write the explicit expression for $p \cdot U$ and $p \cdot V$ which both depend only on $y-\eta$ and the transverse coordinates, namely 
\begin{eqnarray}
p \cdot U &=& p_\perp u_0 \cosh(y-\eta) - {\vec p}_\perp \cdot {\vec u}_\perp , \nonumber \\
p \cdot V &=& p_\perp u_0 \sinh(y-\eta) .
\label{pdotUV}
\end{eqnarray}
From Eqs. (\ref{pdotUV}) one can also infer that in the rest frame system of the plasma element, where $\eta = 0$ and $ {\vec u}_\perp = 0$, we have $p \cdot U = p_\perp \cosh y$ and $p \cdot V = p_\perp \sinh y$. Thus, in the local rest frame of the plasma element Eq. (\ref{Fxp2}) is reduced to Eq. (\ref{Fxp1}).

\subsection{Moments of anisotropic distribution}
\label{sect:aniso-moments}

Using the standard definitions of $N^\mu$ and $T^{\mu \nu}$ as the first and the second moment of the distribution function (\ref{Fxp2}), namely 
\begin{eqnarray}
N^\mu &=& \int \frac{d^3p}{(2\pi)^3 \, E_p} \, p^{\mu} f,
\label{Nmu}
\end{eqnarray}
%
%
\begin{eqnarray}
T^{\mu \nu} &=& \int \frac{d^3p}{(2\pi)^3 \, E_p} \, p^{\mu} p^\nu f,
\label{Tmunu} 
\end{eqnarray}
we obtain the following decompositions:
\begin{eqnarray}
N^\mu &=&  n \, U^\mu,
\label{Nmudec}
\end{eqnarray}
%
%
\begin{eqnarray}
T^{\mu \nu} &=& \left( \varepsilon  + P_T\right) U^{\mu}U^{\nu} 
- P_T \, g^{\mu\nu} - (P_T - P_L) V^{\mu}V^{\nu}. \nonumber \\
\label{Tmunudec}
\end{eqnarray}
We note that $N^\mu$ does not have the contribution proportional to the four-vector $V^\mu$ since such a term would be proportional to the scalar product $V^\mu N_\mu$ that vanishes in the local rest frame. Similarly, the energy-momentum tensor does not contain the terms proportional to the symmetric combination $V^\mu U^\nu + U^\mu V^\nu$, see Ref. \cite{Ryblewski:2008fx} for a more explicit presentation of the analogous decompositions.  

Equation (\ref{Nmudec}) defines the particle density $n$, which may be calculated from the formula
\begin{eqnarray}
n &=& \int \frac{d^3p}{(2\pi)^3} \, f\left( \frac{p_\perp}{\lambda _\perp}, 
\frac{| p_\parallel |}{\lambda _\parallel }\right) \label{rho1} \\
&=& \frac{\lambda_\perp^2 \,\lambda_\parallel}{2 \pi^2} \int\limits_0^\infty  
d\xi_\perp\,\xi_\perp\, \int\limits_{0}^\infty d\xi_\parallel \, f\left(\xi_\perp,\xi_\parallel \right), \nonumber
\end{eqnarray}
where we have introduced the dimensionless variables 
\begin{equation}
\xi_\perp = \frac{p_\perp}{\lambda_\perp}, \quad \xi_\parallel = \frac{p_\parallel}{\lambda_\parallel}.
\label{xis}
\end{equation}
In the similar way we calculate the energy density,
\begin{eqnarray}
\varepsilon &=&   \int \frac{d^3p}{ (2\pi)^3}  \, E_p \, f\left( \frac{p_\perp}{\lambda _\perp}, \frac{| p_\parallel |}{\lambda _\parallel }\right)  \label{epsilon1} \\
&=& \frac{\lambda_\perp^2 \,\lambda_\parallel^2}{2\pi^2} \int\limits_0^\infty  
d\xi_\perp\,\xi_\perp  \int\limits_0^\infty \,d\xi_\parallel\,
\sqrt{\xi_\parallel^2 + x\, \xi_\perp^2 }
\, f\left(\xi_\perp,\xi_\parallel \right),
\nonumber 
\end{eqnarray}
where the variable $x$ is defined by the expression
\begin{equation}
x = \left( \frac{\lambda_\perp}{\lambda_\parallel} \right)^2.
\label{iks}
\end{equation}
Finally, the transverse and longitudinal pressure is obtained from the equations 
\begin{eqnarray}
P_T &=&   \int \frac{d^3p}{ (2\pi)^3}  \, \frac{p_\perp^2}{2 E_p} \, f\left( \frac{p_\perp}{\lambda _\perp}, \frac{| p_\parallel |}{\lambda _\parallel }\right)  \label{PT1} \\
&=& \frac{\lambda_\perp^4}{2\pi^2} \int  
\frac{d\xi_\perp\,\xi_\perp^3 \,d\xi_\parallel}
{2 \sqrt{ \xi_\parallel^2 + x\, \xi_\perp^2 }}
\, f\left(\xi_\perp,\xi_\parallel \right),
\nonumber 
\end{eqnarray}
\begin{eqnarray}
P_L &=&   \int \frac{d^3p}{ (2\pi)^3}  \, \frac{p_\parallel}{E_p} \, f\left( \frac{p_\perp}{\lambda _\perp}, \frac{| p_\parallel |}{\lambda _\parallel }\right)  \label{PL1} \\
&=& \frac{\lambda_\perp^2 \,\lambda_\parallel^2}{2\pi^2} \int  
\frac{d\xi_\perp\,\xi_\perp \,d\xi_\parallel \,\xi_\parallel^2}
{\sqrt{\xi_\parallel^2 + x\, \xi_\perp^2 }}
\, f\left(\xi_\perp,\xi_\parallel \right).
\nonumber 
\end{eqnarray}
From now on the limits of the integrations over $\xi_\perp$ and $\xi_\parallel$ are always from 0 to infinity. In the local rest-frame of the fluid element, where we have $U^\mu = (1,0,0,0)$ and $V^\mu = (0,0,0,1)$ one finds
\begin{equation}
T^{\mu \nu} =  \left(
\begin{array}{cccc}
\varepsilon & 0 & 0 & 0 \\
0 & P_T & 0 & 0 \\
0 & 0 & P_T & 0 \\
0 & 0 & 0 & P_L
\end{array} \right),
\label{Tmunuarray}
\end{equation}
hence, as expected the structure (\ref{Fxp2}) allows for different pressures in the longitudinal and transverse directions. 

One may also calculate the entropy current using the Boltzmann definition,\footnote{The formula (\ref{S}) assumes the classical Boltzmann statistics. It may be generalized to the case of bosons or fermions in the standard way.}
\begin{equation}
S^{\mu} = g_0\int \frac{d^3p}{(2 \pi)^3} \frac{p^\mu}{ E_p}  \left(\frac{f}{g_0}\right)
 \, \left[1 -  \ln \left(\frac{f}{g_0}\right) \right],
\label{S}
\end{equation}
here $g_0$ is the degeneracy factor related to internal quantum numbers such as spin or color. The entropy current has the structure
\begin{equation}
S^{\mu} = \sigma \, U^\mu,
\label{Sstr}
\end{equation}
where
\begin{eqnarray}
\sigma = \frac{\lambda_\perp^2 \,\lambda_\parallel}{2\pi^2} \int  
d\xi_\perp\,\xi_\perp\,d\xi_\parallel \, f\left(\xi_\perp,\xi_\parallel \right) \left[1 - \ln \frac{f\left(\xi_\perp,\xi_\parallel \right)}{g_0} \right]. \nonumber \\
\label{sigma}
\end{eqnarray}
Comparison of Eqs. (\ref{rho1}) and (\ref{sigma}) indicates that the particle density and the entropy density are proportional, with the proportionality constant depending on the specific choice of the parton distribution function $f$.

\subsection{Pressure relaxation function}
\label{sect:rel-funct}

With the help of the variables $x = (\lambda_\perp/\lambda_\parallel)^2$ and $n$ we may rewrite our expressions (\ref{epsilon1}), (\ref{PT1}), and  (\ref{PL1}) in the concise form
\begin{eqnarray}
\varepsilon &=&  \left(\frac{n}{g} \right)^{4/3} R(x),
\label{epsilon2} 
\end{eqnarray}
\begin{eqnarray}
P_T &=&  \left(\frac{n}{g} \right)^{4/3}
\left[\frac{R(x)}{3} + x R^\prime(x) \right],   
\label{PT2} 
\end{eqnarray}
\begin{eqnarray}
P_L &=&   \left(\frac{n}{g} \right)^{4/3} 
\left[\frac{R(x)}{3} - 2 x R^\prime(x) \right],
\label{PL2} 
\end{eqnarray}
where the function $R(x)$ is defined by the integral
\begin{equation}
R(x) = x^{-1/3} \int \frac{d\xi_\perp\,\xi_\perp\,d\xi_\parallel}{2\pi^2}
\sqrt{\xi_\parallel^2 + x \xi_\perp^2} f(\xi_\perp,\xi_\parallel),
\label{Rofiks}
\end{equation}
$R^\prime(x) = dR(x)/dx$, and $g$ is a constant defined by the expression
\begin{equation}
g =  \int \frac{d\xi_\perp\,\xi_\perp\,d\xi_\parallel}{2\pi^2}
 f(\xi_\perp,\xi_\parallel).
\label{gconst}
\end{equation}

It is quite interesting to observe that the structure of Eqs. (\ref{epsilon2}) - (\ref{PL2}) agrees with the structure derived in \cite{Florkowski:2008ag}, where no reference to the underlying microscopic picture was made but only the general consistency of the approach based on the anisotropic energy-momentum tensor (\ref{Tmunudec}) and the conservation laws was studied. The only difference is that the entropy density $\sigma$ used in Ref. \cite{Florkowski:2008ag} is now replaced by the particle density $n$. 

In fact, one may repeat the arguments presented in \cite{Florkowski:2008ag} replacing the assumption of the conservation of entropy by the assumption of the particle-number conservation  (note that we have shown above that $n$ and $\sigma$ are proportional if one uses the ansatz (\ref{Fxp1})). In such a case we end up with the structure which exactly matches Eqs. (\ref{epsilon2}) - (\ref{PL2}) and $R$ may be identified with the pressure relaxation function. Moreover, the results of Ref. \cite{Florkowski:2008ag} allow us to relate the variable $x = \lambda_\perp^2/\lambda_\parallel^2$ with the quantity $n \tau^3$ -- we shall come back to the discussion of this point below, after the analysis of some special cases of the anisotropic distribution functions.

\subsection{Boltzmann-like anisotropic distribution}
\label{sect:aBoltz}

As the special case of the anisotropic distribution function we may consider the exponential distribution of the form 
\begin{equation}
f_1 = g_0 \exp \left( -\sqrt{\frac{p_\perp ^2}{\lambda_\perp  ^2} + 
\frac{p_\parallel^2}{\lambda_\parallel^2} } \,  \right),
\label{aBoltz1}
\end{equation}
which may be regarded as the generalization of the Boltzmann equilibrium distribution where $\lambda_\perp = \lambda_\parallel = T$ (as explained above, $g_0$ is the degeneracy factor connected with internal quantum numbers). In this case we recover the structure (\ref{epsilon2}) - (\ref{PL2}) with the relaxation function of the form \footnote{Note that for $x < 1$ the function $(\arctan\sqrt{x-1})/\sqrt{x-1}$ should be replaced by $(\hbox{arctanh}\sqrt{1-x})/\sqrt{1-x}$}
\begin{equation}
R_1(x) = \frac{3\, g_0\, x^{-\frac{1}{3}}}{2 \pi^2} \left[ 1 + \frac{x \arctan\sqrt{x-1}}{\sqrt{x-1}}\right]
\end{equation}
and the constant (\ref{gconst}) is simply 
\begin{equation}
g_1 = \frac{g_0}{\pi^2}.
\label{g1const}
\end{equation}

Another interesting anisotropic distribution function has the factorized form
\begin{equation}
f_2 = g_0 \exp\left( -\frac{p_\perp}{\lambda_\perp} \right)
\exp\left( - \frac{|p_\parallel|}{\lambda_\parallel} \right).
\label{aBoltz2}
\end{equation}
In this case we obtain
\begin{eqnarray}
R_2(x) &=& \frac{g_0 x^{-1/3}}{2 \pi^2 (1+x)^2} \left[\,
1 + 5 x \sqrt{x} + 2 x^2 \sqrt{x} - 2 x
\vphantom{\frac{1+\sqrt{x}+\sqrt{1+x}}{1+\sqrt{x}-\sqrt{1+x}}}
\right. \nonumber \\
& & \left. \quad + \frac{3 x}{\sqrt{x+1}} 
\ln \frac{1+\sqrt{x}+\sqrt{1+x}}{1+\sqrt{x}-\sqrt{1+x}} \,\,
\right]
\end{eqnarray}
and 
\begin{equation}
g_2 = \frac{g_0}{2 \pi^2}.
\label{g1const}
\end{equation}
The calculation of the entropy density gives $\sigma = 4 n$ and $\sigma = 8 n$, for the cases $f=f_1$ and $f=f_2$, respectively.

The structure of Eq. (\ref{Rofiks}) implies that for $x \ll 1$ \mbox{($\lambda_\perp \ll \lambda_\parallel$)} the function $R(x)$ behaves like $x^{-1/3}$. In this limit $P_T = 0$ and $\varepsilon = P_L$. Similarly, for $x \gg 1$ \mbox{($\lambda_\perp \gg \lambda_\parallel$)} the function $R(x)$ behaves like $x^{1/6}$, implying that $P_L = 0$ and $\varepsilon = 2 P_T$. This behavior is expected if we interpret the parameters $\lambda_\perp$ and $\lambda_\parallel$ as the transverse and longitudinal temperatures, respectively. In agreement with those general properties we find 
\begin{eqnarray}
R_1(x) &\approx &  \frac{3 g_0 }{2 \pi^2} 
\left[ x^{-1/3} + \frac{1}{2}(\ln 4 - \ln x) x^{2/3} \right] , 
\nonumber \\
R_2(x) &\approx & \frac{g_0 }{2 \pi^2} 
\left[ x^{-1/3} +\frac{1 }{2} (6 \ln 2 - 8 - 3 \ln x) x^{2/3}\right] , 
\nonumber \\
\label{smalliks}
\end{eqnarray}
for $x \ll 1$, and
\begin{eqnarray}
R_1(x) &\approx & \frac{3 g_0 }{4 \pi}
\left( x^{1/6} + \frac{1}{2} x^{-5/6}\right),
\nonumber \\
R_2(x) &\approx & \frac{g_0 }{\pi^2}
\left( x^{1/6} + \frac{1}{2} x^{-5/6}\right),
\label{bigiks}
\end{eqnarray}
for $x \gg 1$. 

\begin{figure}[t]
\begin{center}
\subfigure{\includegraphics[angle=0,width=0.45\textwidth]{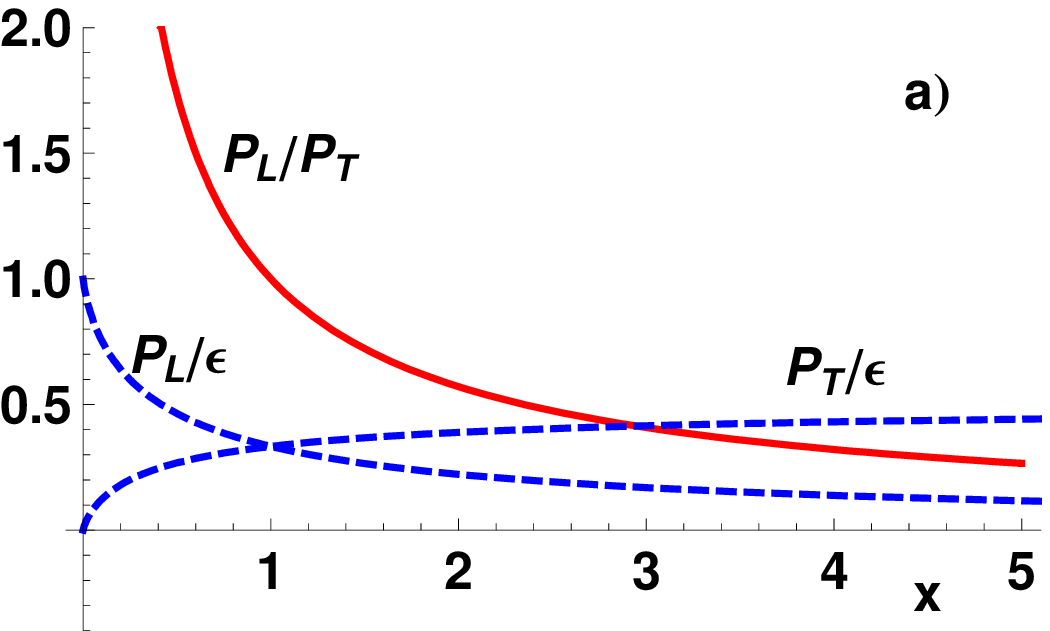}} \\
\subfigure{\includegraphics[angle=0,width=0.45\textwidth]{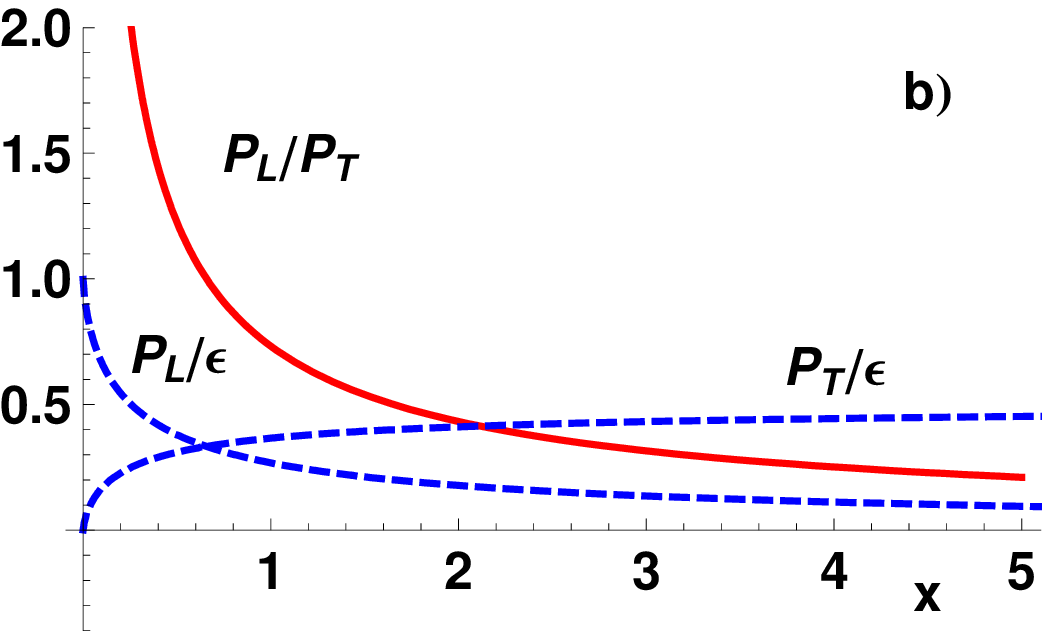}} 
\end{center}
\caption{(Color online) The ratios $P_L/P_T$ (solid red lines), $P_L/\varepsilon$ (decreasing blue dashed lines), and $P_T/\varepsilon$ (increasing blue dashed lines) shown as functions of the variable $x$, {\bf a)} the results for the distribution function (\ref{aBoltz1}), {\bf b)} the same for the distribution function (\ref{aBoltz2}). }
\label{fig:ratios}
\end{figure}

In Fig. \ref{fig:ratios} we plot the ratios: $P_L/P_T$ (solid line), $P_L/\varepsilon$ (decreasing dashed line), and $P_T/\varepsilon$ (increasing dashed line) for the two cases: $f = f_1$ (a) and $f=f_2$ (b). The considered ratios are functions of the $x$ parameter only. In agreement with the remarks given above we see that $\varepsilon = P_L$ for $x = 0$, and $\varepsilon = 2 P_T$ in the limit $x \to \infty$. For $f = f_1$ the two pressures become equal if  $x=1$, since in this case the distribution function $f_1$ becomes exactly isotropic. For $f = f_2$ the equality of pressures is reached for $x \approx 0.7$. Except for such small quantitative differences, the behavior of the pressures is very much similar in the two cases, as can be seen from the comparison of the upper and lower part of Fig. \ref{fig:ratios}.

We note that the choice $R(x) = x^{1/6}$ corresponds to the case of transverse hydrodynamics, see Ref. \cite{Florkowski:2008ag}. In the transverse-hydrodynamic approach the matter forms non-interacting transverse clusters which do not interact with each other yielding $P_L=0$. The concept of transverse hydrodynamics was initiated in Refs. \cite{Heinz:2002rs,Heinz:2002xf} and recently reformulated in Refs. \cite{Bialas:2007gn,Chojnacki:2007fi,Ryblewski:2008fx}.

\subsection{Time dependence of pressure anisotropy}
\label{sect:PToverPL}

In this Section we briefly recall the arguments of Ref. \cite{Florkowski:2008ag} concerning the consistency of the anisotropic plasma dynamics. The basic assumptions are the particle number conservation,
\begin{equation}
\partial_\mu N^\mu = \partial \left( n U^\mu \right) = 0,
\label{partcons}
\end{equation}
and the energy-momentum conservation law,
\begin{equation}
\partial_\mu T^{\mu \nu} = 0,
\label{enmomcon}
\end{equation}
with the energy-momentum tensor of the form (\ref{Tmunudec}). As shown in the previous Sections, the entropy conservation used in \cite{Florkowski:2008ag} may be typically identified with the particle number conservation. In view of the further development of the model discussed in the next Section we shall turn to the particle number conservation as the basic input. We note that the assumption (\ref{partcons}) means that our description may be valid only after the time when most of the particles is produced. 

The projection of the energy-momentum conservation law (\ref{enmomcon}) on the four-velocity $U_\nu$ indicates that the energy density is generally a function of two variables, \mbox{$\varepsilon = \varepsilon(n,\tau)$}. The mathematical consistency of this approach, i.e., the requirement that $d\varepsilon$ is a total differential, implies directly that the functions $\varepsilon(n,\tau)$, $P_T(n,\tau)$, and $P_L(n,\tau)$ must be of the form (\ref{epsilon2}) - ({\ref{PL2}) where
\begin{equation}
x = x_0 \frac{n \tau^3}{n_0 \tau_0^3}
\label{oldiks}
\end{equation}
with $x_0, n_0$ and $\tau_0$ being constants that may be used to fix the initial conditions. In particular, it is convenient to regard $\tau_0$ as the initial time, and $n_0$ as the maximal initial density (at the very center of the system). Then, $x_0$ is the maximal initial value of $x$. Note that the particle density $n$ is very small at the edge of the system, hence the initial transverse pressure is always zero in this region. On the other hand, at the center of the system at the initial time $\tau=\tau_0$ we may have $P_T < P_L$ or $P_T > P_L$ depending on the value of $x_0$.  

Combing Eqs. (\ref{iks}) and (\ref{oldiks}) we are coming to the main conclusion reached so far: For the microscopic phase-space distribution function of the form (\ref{Fxp1}), the pressure relaxation function is completely determined by Eq. (\ref{Rofiks}) where 
\begin{equation}
x = \frac{\lambda_\perp^2}{\lambda_\parallel^2} =  x_0 \frac{n \tau^3}{n_0 \tau_0^3}.
\label{newiks}
\end{equation}
In the region where the matter is initially formed we have $0 < n \leq n_0$ and the right-hand-side of Eq. (\ref{newiks}) grows with time -- the particle density $n$ cannot decrease faster than $1/\tau^3$, since this would require a three-dimensional expansion at the speed of light. We thus conclude that the ratio of the parameters $\lambda_\parallel/\lambda_\perp$ tends asymptotically in time to zero. Consequently, for sufficiently large evolution times the ratio of the longitudinal and transverse pressures becomes negligible. As mentioned above, even if the initial conditions require that $P_T$ is larger than $P_L$ at the center of the system, at the edges we have very small density which means that $P_L >> P_T$ in this region. In the case where the longitudinal expansion dominates, $n = n_0 \tau_0/\tau$ and $x = x_0 \tau^2/\tau_0^2$, hence $x$ and $\tau$ are simply related.

\subsection{Longitudinal free-streaming}
\label{sect:PToverPL}

The anisotropic distribution functions considered in the previous Sections should satisfy the Boltzmann kinetic equation (in some reasonable approximation). In this respect we assume that the effects of both the free-streaming and the parton collisions do not change the generic structure (\ref{Fxp2}), while the time changes of the parameters $\lambda_\perp$,$\lambda_\parallel$, and $u^\mu$ are determined by the conservation laws. The spirit of this approach is very similar to that used in the perfect-fluid hydrodynamics, where the collisions maintain the equilibrium shape of the distribution function, whereas the conservation laws determine the time changes of the parameters such as temperature or the fluid velocity. 

Clearly the relation of our framework to the underlying kinetic theory should be elaborated in more detail in further investigations, which may possibly determine the microscopic conditions which validate our approximations. Here, we may easily analyze the case of pure free-streaming where the distribution function satisfies the collisionless kinetic equation
\begin{equation}
p^\mu \partial_\mu f(x,p) = 0.
\label{kineq}
\end{equation}
For the pure longitudinal expansion (with vanishing transverse flow, ${\vec u}_\perp=0$, and the parameters $\lambda_\perp,\lambda_\parallel$ depending only on the proper time $\tau$) we rewrite Eq. (\ref{kineq}) in the form 
\begin{equation}
\left[ \cosh(y-\eta) \frac{\partial}{\partial\tau}
+ \frac{\sinh(y-\eta)}{\tau} \frac{\partial}{\partial \eta} \right] f\left(w,v\right) = 0,
\label{kineq1}
\end{equation}
where $w = p_\perp/\lambda_\perp(\tau)$ and $v = p_\perp \sinh(y-\eta)/\lambda_\parallel(\tau)$, see Eqs. (\ref{Fxp2}) and (\ref{pdotUV}). By direct differentiation we obtain \footnote{For simplicity we consider here the functions depending on $v^2$ and disregard the sign of the absolute value.}
\begin{equation}
\frac{\partial f}{\partial w} \frac{d\lambda_\perp}{\lambda_\perp^2 d\tau}
+ \frac{\sinh(y-\eta)}{\lambda_\parallel^2} \frac{\partial f}{\partial v}
\left[\frac{d\lambda_\parallel}{d\tau} + \frac{\lambda_\parallel}{\tau}  \right] =0.
\end{equation}
The solution to this equation exists for any form of the function $f$ provided $\lambda_\perp$ is a constant and $\lambda_\parallel \sim 1/\tau$. Thus, we may write  
\begin{equation}
f = f\left( \frac{p_\perp}{\lambda_\perp^0}, \frac{\tau p_\perp \sinh(y-\eta)}{\tau_0 \lambda_\parallel^0} \right),
\label{freestreamsol}
\end{equation}
where $\tau_0$, $\lambda_\perp^0$ and $\lambda_\parallel^0$ are constants. 

In the considered case the variable $x$ equals
\begin{equation}
x = \left( \frac{\lambda_\perp}{\lambda_\parallel} \right)^2 =
\left( \frac{\lambda_\perp^0 }{\lambda_\parallel^0 \tau_0} \right)^2 \, \tau^2
\end{equation}
hence, it is consistent with Eq. (\ref{newiks}), where for the boost-invariant longitudinal expansion we may substitute $n = n_0 \tau_0/\tau$. We thus see that our approach includes the boost-invariant free-streaming as the special case. In particular, Eq. (\ref{freestreamsol}) agrees with the form of the color neutral background used in Refs. \cite{Rebhan:2008uj,Rebhan:2008ry,Martinez:2008di}. In the next Section we show that our framework includes also the case where the partons interact with local magnetic fields. 

\section{Locally anisotropic magnetohydrodynamics}
\label{sect:AMHD}

In this Section we generalize the formulation discussed in Sect. \ref{sect:aniso-system}. We analyze in detail magnetohydrodynamics as an example of the physical system consisting of particles and fields, which is also known to exhibit strong anisotropic behavior. Of course, the magnetohydrodynamics by itself cannot be directly applied to modeling of the early stages of heavy-ion collisions. However, several phenomena analyzed in its framework show similarities with the color field dynamics discussed in the context of Color Glass Condensate \cite{McLerran:1993ni,Kharzeev:2001yq} and Glasma \cite{Lappi:2006fp}, hence we think that the elaboration of this example may shed light on more complicated color-hydrodynamics which may be the right description of the early stages of heavy-ion collisions. 

Our analysis of the boost-invariant magnetohydrodynamics, where the initial magnetic field is parallel to the collision axis,  shows that in the considered system a similar phenomena take place as in the system consisting of particles only. The presence of the fields lowers the longitudinal pressure (which eventually may be negative) and increases the transverse pressure, see a related analysis in \cite{Vredevoogd:2008id}.

\subsection{General formulation}
\label{sect:binv}

At first let us recapitulate the main physical assumptions of locally anisotropic magnetohydrodynamics (for recent formulation see for example \cite{PhysRevE.47.4354,PhysRevE.51.4901}). Let $U^\mu$ be the plasma four-velocity and $F^{\mu \nu}$ be the electromagnetic-field tensor. We define the rest-frame electric and magnetic field by the following equations 
\begin{equation}
E^\mu = F^{\mu \nu} U_\nu, 
\label{Emu}
\end{equation}
\begin{equation}
B^\mu = \frac{1}{2} \epsilon^{\mu \alpha \beta \gamma} U_\alpha F_{\beta \gamma},
\label{Bmu}
\end{equation}
where $\epsilon^{\alpha \beta \gamma \delta}$ is a completely antisymmetric tensor with $\epsilon^{0123} = 1$. Eqs. (\ref{Emu}) and (\ref{Bmu}) yield
\begin{equation}
F^{\mu \nu} = E^\mu U^\nu - E^\nu U^\mu + \frac{1}{2} \epsilon^{\mu \nu \alpha \beta} \left(B_\alpha U_\beta - B_\beta U_\alpha \right).
\label{Fmunu1}
\end{equation}
We note that both $E^\mu$ and $B^\mu$ are spacelike and orthogonal to $U^\mu$,
\begin{eqnarray}
E_\mu E^\mu & \leq & 0, \quad E_\mu U^\mu = 0, \label{EU} \\
B_\mu B^\mu & \leq & 0, \quad B_\mu U^\mu = 0. 
\label{BU} 
\end{eqnarray}

The picture  of anisotropic magnetohydrodynamics requires that $U^\mu$ corresponds to the frame where the electric field is absent, 
\begin{equation}
E^\mu = 0.
\label{Emu0}
\end{equation}
In this case the Maxwell equations may be written in the form
\begin{equation}
\partial_\mu F^{\mu \nu} = 4 \pi j^\nu, 
\label{maxwell1}
\end{equation}
\begin{equation}
\partial_\mu {}^* F^{\mu \nu} = 0,
\label{maxwell2}
\end{equation}
where
\begin{equation}
F^{\mu \nu} = \frac{1}{2} \epsilon^{\mu \nu \alpha \beta} (B_\alpha U_\beta - B_\beta U_\alpha)
\label{Fmunu2}
\end{equation}
and ${}^* F^{\mu \nu}$ is the dual electromagnetic tensor
\begin{equation}
{}^*F^{\mu \nu} =  B^\mu U^\nu - B^\nu U^\mu.
\label{Fdual}
\end{equation}
Besides Eqs. (\ref{maxwell1}) -- (\ref{Fdual}) the plasma dynamics is determined by the particle conservation law, see Eq. (\ref{partcons}), the electromagnetic current conservation (the consequence of Eq. (\ref{maxwell1})), and the energy-momentum conservation law for matter and fields. Before we analyze the role of the conservation laws we shall discuss, however, the constraints coming from the boost-invariance. 

\subsection{Imposing longitudinal boost-invariance}
\label{sect:binv}

Our aim is to construct the boost-invariant field tensors $F^{\mu \nu}(x)$ and ${}^*F^{\mu \nu}(x)$. The condition of boost-invariance requires that the transformed fields at the new spacetime positions are equal to the original fields in the new positions. Formally, this condition may be written in the form
\begin{equation}
F^{\mu \nu \, \prime}(x^\prime) = L^\mu_{\,\,\,\alpha} L^\nu_{\,\,\,\beta }   
F^{\alpha \beta}(x) = F^{\mu \nu}(x^\prime).
\end{equation}
where $L$ describes the longitudinal Lorentz boost. Similarly, for a boost-invariant four-vector field $A^\mu(x)$ we have
\begin{equation}
A^{\mu \,\prime}(x^\prime) = L^\mu_{\,\,\,\alpha} 
A^{\alpha}(x) = A^{\mu}(x^\prime).
\end{equation}
It is easy to check that the four-vectors $U^\mu$ and $V^\mu$ defined by Eqs. (\ref{U}) and (\ref{V}) are invariant under Lorentz boosts with rapidity $\alpha$ along the longitudinal axis, defined by the matrix
\begin{equation}
L^{\mu}_{\,\, \nu}(\alpha) =
\left(
\begin{array}{cccc}
\cosh \alpha & 0 & 0 & \sinh \alpha \\
0 & 1 & 0 & 0 \\
0 & 0 & 1 & 0 \\
\sinh \alpha & 0 & 0 & \cosh \alpha
\end{array} 
\right).
\label{Lmunu}
\end{equation}
Since $U^\mu$ and $V^\mu$ are boost-invariant, the structure of Eqs. (\ref{Fmunu2}) and (\ref{Fdual}) suggests that the boost-invariant formalism follows from the ansatz 
\begin{equation}
B^\mu = B V^\mu.
\label{BmuVmu}
\end{equation}
where $B$ is a scalar function depending on $\tau$ and transverse coordinates ${\vec x}_\perp$. Equation (\ref{BmuVmu}) defines the field tensors as the tensor products of the boost-invariant four-vectors, hence, by construction the field tensors are boost-invariant. In addition, we observe that in this case Eq. (\ref{BU}) is automatically fulfilled. 

\subsection{Homogeneous dual field equations}
\label{sect:conlaw}

Projection of the homogeneous dual field equations (\ref{maxwell2}) on the four-velocity $U_\nu$ gives
\begin{equation}
V^\mu \partial_\mu B + B \partial_\mu V^\mu  - B U_\nu U^\mu \partial_\mu V^\nu = 0.
\label{dualeq1}
\end{equation}
For the boost-invariant systems all terms in (\ref{dualeq1}) are identically zero, hence it is automatically fulfilled. On the other hand, the projection of (\ref{maxwell2}) on the four-vector $V_\nu$ gives
\begin{equation}
U^\mu \partial_\mu \ln \left( \frac{n \tau}{B} \right) = 0,
\label{dualeq2}
\end{equation}
hence $B$ is related to the particle density $n$ and the proper time $\tau$ by the expression
\begin{equation}
B = B_0 \frac{n \tau}{n_0 \tau_0}.
\label{Bntau}
\end{equation}
One may check that with the ansatz (\ref{Bntau}) all four equations in (\ref{maxwell2}) are automatically satisfied, hence Eq. (\ref{Bntau}) is the main piece of information delivered by the homogeneous dual field equations. In particular the equation ${\vec \nabla} \cdot {\vec B} = 0$ turns out to be equivalent with the continuity equation for the particle number. 

Collecting now Eqs. (\ref{BmuVmu}) and (\ref{Bntau}) we find the explicit form of the dual field tensor ${}^* F^{\mu \nu} $,
\begin{widetext}
\begin{eqnarray}
{}^* F^{\mu \nu} =  \frac{B_0 n \tau}{n_0 \tau_0} \left(
\begin{array}{cccc}
0              & u_x \sinh\eta  & u_y \sinh\eta  & -u^0 \\
-u_x \sinh\eta & 0             & 0 & -u_x \cosh\eta  \\
-u_y \sinh\eta & 0             & 0 & -u_y \cosh\eta  \\
u^0 & u_x \cosh\eta  & u_y \cosh\eta  & 0
\end{array} \right). \nonumber \\
\end{eqnarray}
\end{widetext}

The structure of the dual tensor allows us to infer the form of the electric and magnetic fields, 
\begin{eqnarray}
{\vec B} &=&  \frac{B_0 n \tau}{n_0 \tau_0}\left(- u_x \sinh\eta  , -u_y \sinh\eta  , u^0  \right),
\label{vecB1} \\
{\vec E} &=&  \frac{B_0 n \tau}{n_0 \tau_0}\left(- u_y \cosh\eta  , u_x \cosh\eta  , 0  \right),
\label{vecE1}
\end{eqnarray}
The above structure implies directly that with no transverse expansion, i.e., for $u_x=u_y=0$ only the longitudinal magnetic field is present in the system and, in view of the relation $n=n_0 \tau_0/\tau$, it should be a constant, $B=B_0$.  

In our general approach the situation $u_x=u_y=0$ corresponds  to the initial condition for the evolution of the system. It resembles the case of the Glasma \cite{Lappi:2006fp} where also the longitudinal chromo-magnetic field is present, however, in the case of Glasma the direction of the field is random with the coherence transverse length set by the saturation scale (another difference is the presence of the longitudinal chromo-electric field in the Glasma). When the transverse expansion starts, due to the presence of the transverse pressure, it initiates the formation of the transverse magnetic and electric fields which are always perpendicular to each other, ${\vec B} \cdot {\vec E} = 0$. We note, however, that in the local rest frame of the plasma element, the only non-zero component is $B_z$.

A more compact form representing the fields ${\vec B}$ and ${\vec E}$ may be achieved if we use the following parameterization of the particle current 
\begin{eqnarray}
N^\mu &=& n \left(u^0 \cosh\eta, u_x, u_y, u^0 \sinh\eta \right) \nonumber \\
&=& \left(n \, u^0 \cosh\eta, n_x, n_y, n u^0 \sinh\eta \right).
\label{Nmun}
\end{eqnarray}
Using the quantities $n_x$ and $n_y$ we write
\begin{eqnarray}
{\vec B} &=&  \frac{B_0}{n_0 \tau_0}\left(- z\, n_x   , - z\, n_y  , n\, \tau\, u^0  \right),
\label{vecB2} \\
{\vec E} &=&  \frac{B_0}{n_0 \tau_0}\left(- t\, n_y   , t\, n_x   , 0  \right).
\label{vecE2}
\end{eqnarray}

\subsection{Inhomogeneous field equations}
\label{sect:conlaw}

We turn now to the inhomogeneous field equations (\ref{maxwell1}). In our approach those equations may be used to determine  the electromagnetic current of the system, $j^\mu = (\rho, j_x, j_y, j_z)$. The straightforward calculation, where the form of the magnetic and electric fields given by Eqs.  (\ref{vecB2}) and (\ref{vecE2}) is used, leads us to the expressions 
\begin{eqnarray}
j^0 = \rho &=& \frac{B_0\,t}{n_0 \tau_0} \left(\partial_y n_x - \partial_x n_y \right), \nonumber \\
j^1 = j_x &=&  \frac{B_0}{n_0 \tau_0} \left[\tau \partial_y (n u^0) + 2 n_y + \tau \partial_\tau n_y \right], \nonumber \\
j^2 = j_y &=&  \frac{B_0}{n_0 \tau_0} \left[-\tau \partial_x (n u^0) - 2 n_x - \tau \partial_\tau n_x \right], \nonumber \\
j^3 = j_z &=&  \frac{B_0\,z}{n_0 \tau_0} \left(\partial_y n_x - \partial_x n_y \right).
\label{jmu}
\end{eqnarray}
One may check by the explicit calculation that the electromagnetic four-current $j^\mu$ defined by Eq. (\ref{jmu}) is conserved, as required by the equation (\ref{maxwell1}). 

In the magnetohydrodynamic appoach one usually assumes that matter is neutral. In our case, the neutrality condition $\rho = 0$ implies that the flow must be rotationless, i.e.,  the following equation should be satisfied 
\begin{equation}
\partial_y n_x - \partial_x n_y = 0.
\label{rotless}
\end{equation}
In this case also the longitudinal component of the electromagnetic current vanishes, which means that the non-zero current circulates around the $z$-axis. 

The explicit calculation with the magnetic and electric fields given by Eqs.  (\ref{vecB2}) and (\ref{vecE2}) shows also that
\begin{equation}
{\vec E} + {\vec v} \times {\vec B} = 0.
\label{Ohm1}
\end{equation}
This is nothing else but the non-covariant version of the condition (\ref{Emu0}).

\subsection{Conservation laws}
\label{sect:conlaw}

Besides the Maxwell equations, the equations of magnetohydrodynamics include the conservation laws for: the particle number, the electromagnetic current (following directly from Eq. (\ref{maxwell1})), and the energy-momentum of the combined system consisting of matter and fields. The total energy-momentum conservation law may be written in the form 
\begin{equation}
\partial_\mu {\hat T}^{\mu \nu} = 0.
\label{enmomconhat}
\end{equation}
where the energy-momentum tensor, ${\hat T}^{\mu \nu}$, including the contributions from matter and fields has the structure
\begin{eqnarray}
{\hat T}^{\mu \nu} &=& \left(\varepsilon + P_T + \frac{B^2}{4\pi} \right) U^\mu U^\nu
-\left(P_T + \frac{B^2}{8\pi} \right) g^{\mu \nu} \nonumber \\
&+& \left(P_L - P_T - \frac{B^2}{4\pi} \right) V^\mu V^\nu,
\label{Thatmunu}
\end{eqnarray}
One may reduce the tensor (\ref{Thatmunu}) to the form (\ref{Tmunudec}) if we introduce the following variables
\begin{eqnarray}
{\hat \varepsilon} &=& \varepsilon + \frac{B^2}{8 \pi} 
= \varepsilon + {\bar \varepsilon}, \nonumber \\
{\hat P_T} &=& P_T + \frac{B^2}{8 \pi} = P_T + {\bar P}_T, \nonumber \\
{\hat P_L} &=& P_L - \frac{B^2}{8 \pi} = P_L + {\bar P}_L.
\label{hatvariables}
\end{eqnarray}
Clearly, the variables with a hat describe the sum of the matter and field contributions to the total energy density and transverse/longitudinal pressures (the field contributions are marked with a bar).

\begin{figure}[t]
\begin{center}
\subfigure{\includegraphics[angle=0,width=0.45\textwidth]{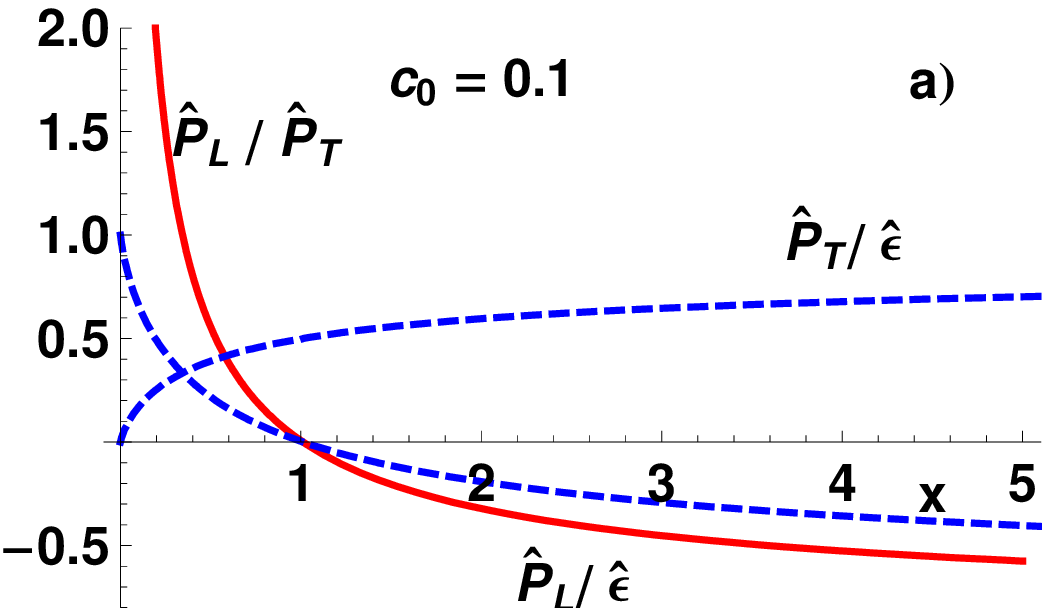}} \\
\subfigure{\includegraphics[angle=0,width=0.45\textwidth]{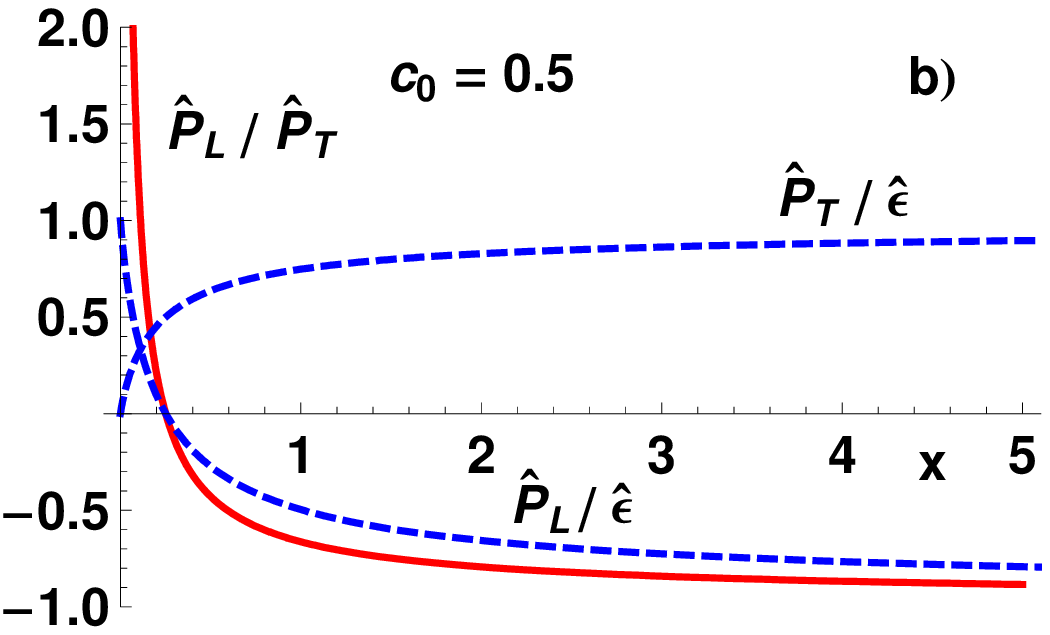}} 
\end{center}
\caption{(Color online) The ratios: ${\hat P}_L/{\hat P}_T$ (solid red lines), ${\hat P}_L/{\hat \varepsilon}$ (decreasing blue dashed lines), and ${\hat P}_T/{\hat \varepsilon}$ (increasing blue dashed lines) shown as functions of the variable $x$, \mbox{{\bf a)} the results} for the distribution function (\ref{aBoltz1}) and $c=0.1$, {\bf b)} the same for $c=0.5$.}
\label{fig:ratiosc}
\end{figure}

Following the same method as that introduced in Ref. \cite{Florkowski:2008ag} we find that the energy-momentum conservation leads to the differential equation
\begin{equation}
d{\hat \varepsilon} = \frac{{\hat \varepsilon}+{\hat P_T} }{n} dn + 
\frac{{\hat P_T} -  {\hat P_L} }{\tau} d\tau.
\label{dhateps}
\end{equation}
Exactly this structure implies that the energy density and pressures are of the form (\ref{epsilon2}) -- (\ref{PL2}). So we may immediately write
\begin{eqnarray}
{\hat \varepsilon} &=&  \left(\frac{n}{g} \right)^{4/3} {\hat R}(x),
\label{epsilon3} 
\end{eqnarray}
\begin{eqnarray}
{\hat P}_T &=&  \left(\frac{n}{g} \right)^{4/3}
\left[\frac{{\hat R}(x)}{3} + x {\hat R}^\prime(x) \right],   
\label{PT3} 
\end{eqnarray}
\begin{eqnarray}
{\hat P}_L &=&   \left(\frac{n}{g} \right)^{4/3} 
\left[\frac{{\hat R}(x)}{3} - 2 x {\hat R}^\prime(x) \right], 
\label{PL3} 
\end{eqnarray}
where the complete relaxation function for matter and fields equals
\begin{equation}
{\hat R}(x) = R(x) + c_0 x^{2/3},
\label{hatR}
\end{equation}
with the parameter $c_0$ defined by the equation
\begin{equation}
c_0 = \frac{B_0^2}{8\pi} \left( \frac{g}{n_0}\right)^{4/3} x_0^{-2/3}.
\label{c0}
\end{equation}

In Fig. \ref{fig:ratiosc} we show the ratios: ${\hat P}_L/{\hat P}_T$ (solid red lines), ${\hat P}_L/{\hat \varepsilon}$ (decreasing blue dashed lines), and ${\hat P}_T/{\hat \varepsilon}$ (increasing blue dashed lines) shown as functions of the variable $x$. Similarly to the case without the magnetic field we observe that the ratio of the longitudinal and transverse pressures decreases with $x$. The new feature of the case with the field is, however, that this ratio may become negative. This behavior reflects the negative contribution of the field pressure $ {\bar P}_L = -B^2/(8\pi)$ to the total pressure ${\hat P}_L$. It becomes dominant for the large values of $x$, where the matter  contribution, growing as $x^{1/6}$, may be neglected with the field contribution, growing as $x^{2/3}$.

In view of our discussion in Sect. II, the variable $x$ depends monotonically on time, hence the $x$ dependence reflects to large extent the time evolution of the studied ratios. If the initial conditions assume very small value of $x_0$ (and consequently of the initial $x$) the system has initially larger total longitudinal pressure than the transverse pressure ${\hat P}_T$. The time evolution tends to equilibrate and then to invert the ratio of the two pressures. The time scale for this process is determined by the initial value of the field, $B_0$, as can be noticed by the comparison of the upper and lower part of Fig. \ref{fig:ratiosc}. 

We close this section with the following remark. Since, $B$ is a function of $n$ and $\tau$, we may rewrite Eq. (\ref{dhateps}) in the equivalent form as
\begin{equation}
d{\varepsilon} = \frac{{\varepsilon}+{P_L} }{n} dn + 
\frac{{P_T} -{P_L} }{B} dB.
\label{dhatepsnew}
\end{equation}
This equation displays the dependence of the energy density $\varepsilon$ on the particle density $n$ and the magnetic field $B$. The equation of the form  $\varepsilon = \varepsilon(n,B)$ plays a role of the equation of state. For the boost-invariant systems the functional dependence $\varepsilon(n,B)$ may be changed to the non-trivial dependence of $\varepsilon$ on $n$ and $\tau$, as introduced in Ref. \cite{Florkowski:2008ag}.

\section{Conclusions}

In this paper we have developed the formalism introduced in Ref. \cite{Florkowski:2008ag} discussing i) the system described by the anisotropic distribution function and ii) the system of partons interacting with local magnetic fields. The presented results may be used to analyze anisotropic systems formed in relativistic heavy-ion collisions. In particular, they may be used to find anisotropic neutral distribution functions which form the background for the field instabilities possibly responsible for the genuine thermalization/isotropization. In addition, our analysis indicates that the process of stable isotropization may require that the assumption concerning boost-invariance and/or entropy conservation should be relaxed.   
 
\medskip
Acknowledgements: We thank W. Broniowski and \mbox{St. Mr\'owczy\'nski} for helpful discussions and critical comments.


\end{document}